\documentclass[aps,prl,10pt,twocolumn,showpacs,preprintnumbers,amsmath,amssymb,superscriptaddress]{revtex4-1}
\usepackage[english]{babel}
\usepackage[dvips]{graphicx}
\usepackage{dcolumn}
\usepackage{bm}

\usepackage{color}

\begin{document}
\title{Expansion of a radial plasma blast shell into an ambient plasma}

\author{M. E. Dieckmann}
\affiliation{Department of Science and Technology (ITN), Link\"oping University, Campus Norrk\"oping, 60174 Norrk\"oping, Sweden}
\author{D. Doria}
\affiliation{School of Mathematics and Physics, Queen's University Belfast, University Road,
Belfast BT7 1NN, UK}
\author{H. Ahmed}
\affiliation{School of Mathematics and Physics, Queen's University Belfast, University Road,
Belfast BT7 1NN, UK}
\author{L. Romagnani}
\affiliation{LULI, Ecole Polytechnique, CNRS, CEA, UPMC, 91128 Palaiseau, France}
\author{G. Sarri}
\affiliation{School of Mathematics and Physics, Queen's University Belfast, University Road,
Belfast BT7 1NN, UK}
\author{D. Folini}
\affiliation{Universit\'e de Lyon, ENS de Lyon, CNRS, Centre de Recherche Astrophysique de Lyon UMR5574, F-69007, Lyon, France}
\author{R. Walder}
\affiliation{Universit\'e de Lyon, ENS de Lyon, CNRS, Centre de Recherche Astrophysique de Lyon UMR5574, F-69007, Lyon, France}
\author{A. Bret}
\affiliation{ETSI Industriales, Universidad de Castilla-La Mancha, 13071 Ciudad Real, Spain}
\affiliation{Instituto de Investigaciones Energ\'eticas y Aplicaciones Industriales, Campus Universitario de Ciudad Real, 13071 Ciudad Real, Spain}
\author{M. Borghesi}
\affiliation{School of Mathematics and Physics, Queen's University Belfast, University Road,
Belfast BT7 1NN, UK}
\email{corresponding author}

\date{\today}
\begin{abstract}
The expansion of a radial blast shell into an ambient plasma is modeled with a particle-in-cell (PIC) simulation. The unmagnetized plasma consists of electrons and protons. The formation and evolution of an electrostatic shock is observed, which is trailed by ion-acoustic solitary waves that grow on the beam of the blast shell ions in the post-shock plasma. In spite of the initially radially symmetric outflow, the solitary waves become twisted and entangled and, hence, they break the radial symmetry of the flow. The waves and their interaction with the shocked ambient ions slows down the blast shell protons and brings the post-shock plasma closer to an equilibrium.
\end{abstract}
\maketitle
The ablation of a solid target by an intense laser pulse yields a dense and hot blast shell \cite{Maksimchuk00}. Collisions between the plasma particles do not frequently occur on the time-scales of interest and the plasma remains far from a thermal equilibrium. The hot and light electrons expand faster than the ions and the charge separation results in an electric field that accelerates the ions of the blast shell. Depending on the duration and intensity of the laser pulse, they can reach speeds of the order of $10^5-10^7$ m/s via this process, often referred to as target normal sheath acceleration (TNSA) \cite{Wilks01,Romagnani05}. Radiation from the target ionizes any residual gas that was present in the experimental vessel prior to the laser shot. This ambient plasma will resist the expansion of the blast shell. 

During the blast shell's free expansion phase its thermal pressure exceeds by far that of the ambient plasma. The blast shell expands in the form of a rarefaction wave \cite{Auer75,Mora09}, which piles up the ambient plasma ahead of it. A forward shock forms between the piled-up ambient plasma and the pristine ambient plasma. This shock is mediated by collective electrostatic forces if no background magnetic field is present \cite{Bardotti70,Forslund70,Forslund71,Eliasson06}.

The forward shock increases the thermal pressure of the ambient plasma by heating and compressing it. An expansion of the radially symmetric blast shell furthermore implies a density profile that decreases rapidly with increasing radius $r$. The pressure of the shock-compressed ambient plasma will become large enough at some $r$ to slow down the blast shell. Laboratory experiments show that in collisionless plasma ion-acoustic solitons (IAS's) \cite{Romagnani08} and electrostatic shocks \cite{Honzawa73,Ahmed13} emerge in the region where the blast shell interacts most effectively with the ambient plasma. 

Here we perform a PIC simulation to test if we can observe these structures. We model for this purpose the expansion of a circular blast shell in a two-dimensional simulation box with the PIC code \textit{EPOCH} \cite{Arber15}. The radial symmetry, the absence of any strong background magnetic field and the usage of reflecting boundary conditions for particles and fields implies that we only need to resolve one quadrant of the expanding blast shell. We model the quadrant defined by $0\le x \le L$ and $0 \le y \le L$, where the side length \textit{L} = 1.2 mm of the simulation box is resolved by 1500 grid cells along each direction. 

A uniformly distributed plasma, which consists of electrons and protons with the number density $n_0 = 3\times 10^{16}\mathrm{cm}^{-3}$, fills the simulation box at the time $t=0$. The electron density $n_0$ is comparable to that in the ambient plasma in laser-plasma experiments. The electrons (protons) are represented by 100 (200) computational particles (CPs) per cell and the electron (proton) temperature is set to $T_0$ = 1 keV ($T_0/10$). The plasma frequency of the ambient medium is $\omega_p \equiv {(n_0e^2/m_e\epsilon_0)}^{1/2}\approx 10^{13}\textrm{s}^{-1}$ ($e$, $m_e$, $\epsilon_0$: elementary charge, electron mass and vacuum permittivity). The blast shell is modeled by superimposing a second plasma on top of the ambient one in the interval $0 \le r \le L/4$, where $r^2=x^2+y^2$. The densities of the electrons and protons of this second plasma are $24n_0$ and the electrons (protons) of this plasma are represented by 400 (800) CPs per cell. The proton temperature of the blast shell plasma matches that of the ambient plasma while the electron temperature is $4.5T_0$. 

We adjust the numerical weights of the CP's that represent the electrons and protons such that the plasma is initially charge-neutral. No net current is present at $t=0$ and we set $\mathbf{E}(x,y)$ and $\mathbf{B}(x,y)$ to zero. The simulation resolves 530 ps by $3\times 10^5$ time steps. 

The simulation provides us with the spatio-temporal distributions of the proton density $n(x,y,t)$ (normalized to $n_0$) and the normalized energy density $E_E(x,y,t)=e^2\epsilon_0(E_x^2(x,y,t)+E_y^2(x,y,t))/(2m_e^2c^2\omega_p^2)$ of the in-plane electric field. We re-sample both distributions in radial coordinates ($r,\alpha$: radius and azimuth angle relative to $y=0$), which gives $n(r,\alpha,t)$ and $E_E (r,\alpha,t)$. 

Thermal diffusion results in a net flow of electrons from a dense into a dilute plasma and the blast shell plasma goes on a positive potential relative to the ambient one. The ambipolar electric field, which sustains the potential difference, accelerates the ambient electrons that enter the blast shell and decelerates the blast shell electrons that escape into the ambient plasma. Two-stream instabilities, which would otherwise develop in the dense plasma \cite{Dieckmann10}, are suppressed by the larger initial temperature of the blast shell electrons. The ambipolar electric field will accelerate protons towards increasing $r$. 

\begin{figure}
\includegraphics[width=1\columnwidth]{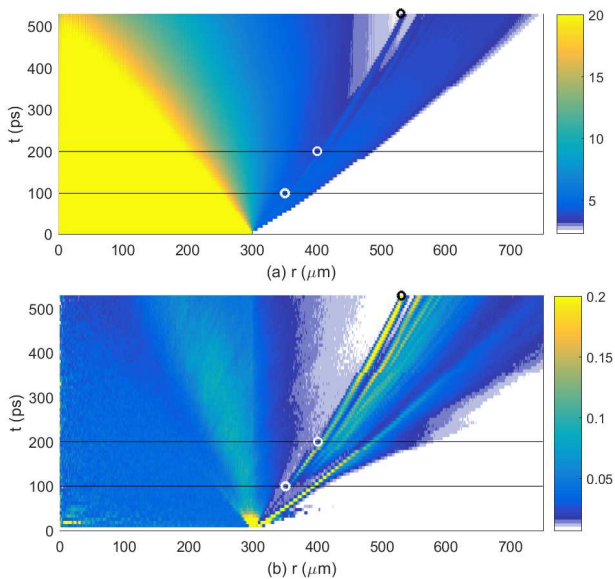}
\caption{Panel (a) shows the averaged proton density distribution $n(r,t)$. The linear color scale is clamped to a minimum of 2.2 and a maximum of 20. Lower and higher densities are thus not resolved by the color map. The averaged energy density $10^3\cdot E_E(r,t)$ of the in-plane electric field is shown in panel (b). The linear color scale is clamped to 0.007 and 0.2. The horizontal lines denote the times $t=100$ ps and $t=200$ ps and the circles are explained in the text.}
\label{expansion}
\end{figure}

Figure \ref{expansion} visualizes this expansion with the help of $n(r,t)$ and $E_E(r,t)$, which are the azimuthal averages of $n(r,\alpha,t)$ and $E_E(r,\alpha,t)$. 
The contour $n(r,t)=20$ in Fig. \ref{expansion}(a) moves in time to lower $r$ and reaches $r=0$ at $t = 530$ ps. Rarefaction waves propagate into the dense plasma and accelerate protons towards the dilute plasma. The density distribution at $t=530$ ps decreases approximately exponentially with increasing $r<400 \, \mu m$. 

A new density bump separates itself from the rarefaction wave at $r\approx 350 \, \mu m$ and $t\approx 100$ ps (lower white circle), which is confined by two boundaries across which the density changes to its maximum $\approx 3$. The right boundary propagates from $r=300 \, \mu m$ at $t=0$ to $r = 750 \, \mu m$ at $t= 530$ ps. Its speed decreases in time and its average is $v_{fs}\approx 8.5 \times 10^5$ m/s. Given that the ion acoustic speed $c_s={(k_B(5T_0/3 + 3T_0/10)/m_p)}$ ($k_B,m_p$: Boltzmann constant and proton mass) of the unperturbed ambient medium is $c_s \approx 4.3 \times 10^5$ m/s, this front is an electrostatic shock with the Mach number $v_{fs}/c_s \approx 2$. A density pulse forms at $t\approx 200$ ps in Fig. \ref{expansion}(a) (upper white circle), which detaches itself from the main bump and reaches $r\approx 530 \mu m$ at $t=530$ ps (black circle). 

A peak of $E_E(r,t)$ forms in Fig. \ref{expansion}(b) at the blast shell boundary $r=300\, \mu m$ immediately after the simulation started. It is the ambipolar electric field that is driven by the initial density jump. Its magnitude exceeds the displayed color range by the factor 10. This initial pulse spreads out and elevated values of $E_E(r,t)$ are present in the interval, which is delimited by the line $r=300 \, \mu m$ and the line that starts at the same position and goes to $r=150\, \mu m$ at $t=530$ ps. This electric field patch outlines the density gradient of the rarefaction wave. 

Statistical fluctuations of the particle number in a volume element in PIC simulation or in real plasma yield fluctuations in the charge- and current density and, hence, electromagnetic fluctuations \cite{Dieckmann04}. The field energy density increases with the particle's thermal energy density. The latter is large in the blast shell plasma, causing elevated level of $E_E(r,t)$ at low $r$ in Fig. \ref{expansion}(b).

The large electric field energy in the density bump at large $r$ is related to the thermalization processes in collisionless plasma. A sharp propagating electric pulse is observed that travels from $300 \, \mu m$ at $t\approx 0$ to $400 \, \mu m$ at $t=100$ ps, after which it starts to become more diffuse. The speed of this pulse is $2.3c_s$ for $0 < t < 100$ ps and it is thus the forward shock. 
 
\begin{figure*}
\includegraphics[width=\textwidth]{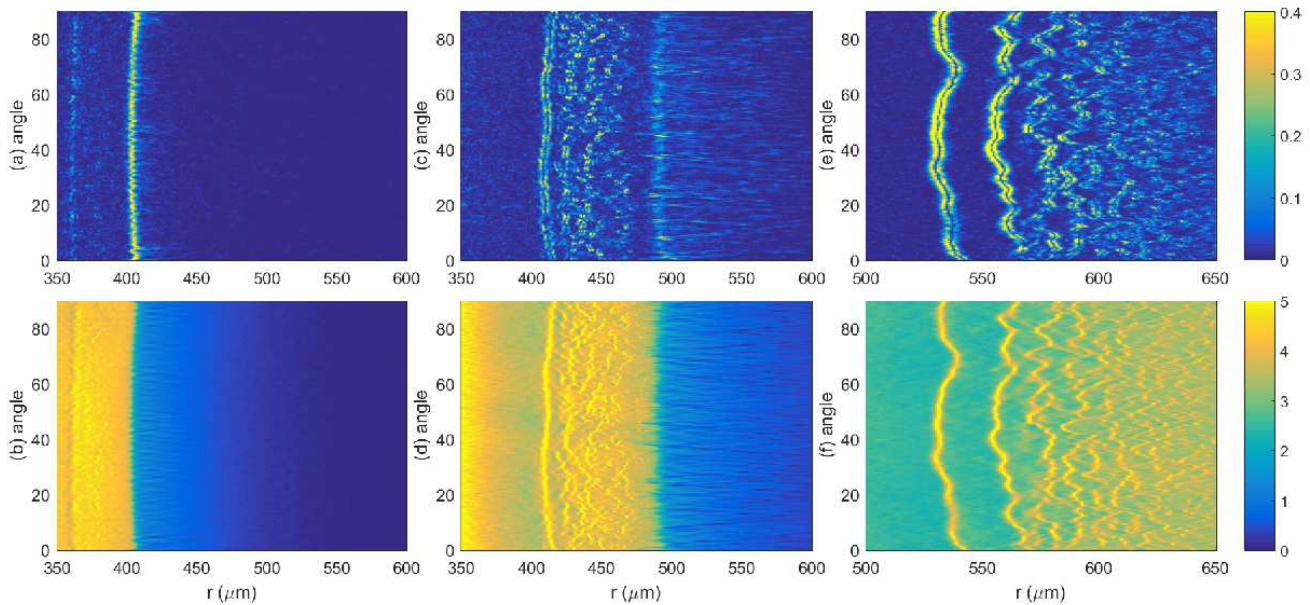}
\caption{The distributions of $E_E(r,\alpha)$ (upper row) and $n(r,\alpha)$ (lower row) at $t=100$ ps (first column), $t=200$ ps (second column) and $t=530$ ps (third column). Both color bars have a linear color scale and are valid for all figures of the respective row (Multimedia view).}
\label{DensityEfield}
\end{figure*}

Figure \ref{DensityEfield} shows the spatial distributions of $n(r,\alpha,t)$ and $E_E(r,\alpha,t)$.
A sharp electric field pulse is present at $r\approx 400\, \mu m$ in Fig. \ref{DensityEfield}(a), which coincides with the density jump between the expanding blast shell and the dilute ambient medium in Fig. \ref{DensityEfield}(b). This pulse is the electrostatic shock. The shock has propagated to $r \approx 490 \, \mu m$ at $t=200$ ps in Fig. \ref{DensityEfield}(c). It has lost its sharpness and waves are observed upstream of it. Such a fragmentation is typical for shocks that reflect a significant part of the inflowing upstream ions, which triggers ion acoustic instabilities \cite{Karimabadi91,Kato10}. The onset of such instabilities explains why the shock has become diffuse in Fig. \ref{expansion}(b). 

Another structure has emerged at $r\approx 410 \, \mu m$ in Fig. \ref{DensityEfield}(c), which is close to the left boundary of the density bump in Fig. \ref{expansion}(a). It reveals two stripes with a large electric field energy density that are separated by a minimum, which coincides with a density spike in Fig. \ref{DensityEfield}(d). We refer with ion solitary wave (ISW) to such a structure. The ISW and the forward shock at $r\approx 490 \, \mu m$ enclose a turbulent region with an elevated plasma density. 

Figures \ref{DensityEfield}(e,f) show $E_E(r,\alpha)$ and $n(r,\alpha)$ at $t=530$ ps. We observe two strong ISWs at $r\approx 540 \, \mu m$ and at $r\approx 570 \, \mu m$ and entangled ones at larger $r$. The average density increases with increasing $r$ in the interval $550\, \mu m \le r \le 650 \, \mu m$, in which we find the entangled ISWs. The trailing ISW had the practically constant value $r=410 \, \mu m$ at $t=200$ ps, while it shows strong variations of $r$ with $\alpha$ at $540 \, \mu m$ and $t=530$ ps. The wavelength of the oscillation at $r\approx 540 \mu m$ and $\alpha \approx 70$ degrees is about 20 degrees, which corresponds for this value of $r$ to an arc length of $\approx 190 \, \mu m$. The amplitude and wave length of this oscillation are close to the values observed at a thin shell of dense ions in a laboratory plasma with similar conditions \cite{Hamad17}.

\begin{figure}
\includegraphics[width=1\columnwidth]{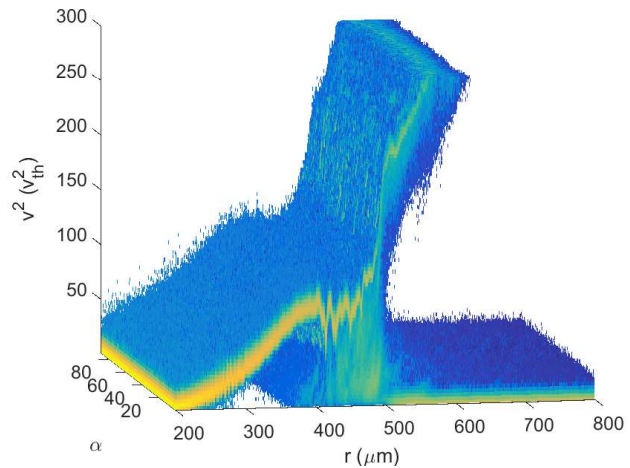}
\caption{The proton phase space density distribution $f_p(r,\alpha,{(|v|/v_{th})}^2)$ at the time 200 ps. The color scale is 10-logarithmic (Multimedia view).}
\label{phasespace}
\end{figure}

The nonlinear plasma structures can be identified unambiguously with the help of the phase space density distribution of the protons. Figure \ref{phasespace} shows  the phase space density distribution $f_p(r,\alpha,{(|v|/v_{th})}^2)$ of the protons at the time 200 ps, where $v_{th}={(k_BT_0/10m_p)}^{1/2}$ is the proton thermal speed $\approx 10^5$ m/s.
The mean velocity of the rarefaction wave increases linearly with $r$ in the interval $300\, \mu m \le r \le 360\, \mu m$ and its velocity increase slows down between $r=360 \, \mu m$ and $r = 400 \, \mu m$. The reduced acceleration is caused by the presence of shocked ambient protons at this location. The density contribution of these protons decreases the proton density gradient in this interval and, hence, the ambipolar electric field that accelerates the protons of the rarefaction wave. 

The mean velocity of the blast shell protons oscillates in the interval $400\, \mu m \le r \le 450 \, \mu m$ with an amplitude that exceeds $v_{th}$ significantly. These velocity oscillations correspond to the previously observed ISWs. They are immersed in hot protons, which originate from the shock-heated ambient protons. They form a dilute cloud with a large thermal spread in the interval $400 \, \mu m \le r \le 500\, \mu m$. A forward shock is located at $r\approx 500 \, \mu m$, which moves to increasing $r$. The ambient protons that cross this shock are heated to the downstream temperature. A fraction of the protons is reflected by the shock potential. These protons feed the dilute low-energy part of the energetic proton beam in the interval $r>500\, \mu m$. Reflected protons are a characteristic of electrostatic shocks. Blast wave protons that crossed the downstream region and were accelerated to larger energies by the shock potential form the denser high-energy part of the fast proton beam ahead of the shock. The shock thus also acts as a double layer \cite{Hershkowitz81}. The interaction between the energetic proton beam and the ambient protons in the interval $r>500\, \mu m$ causes an ion-ion instability \cite{Forslund70b}, which replaces the narrow uni-polar electric field pulse at $r=400\, \mu m$ in Fig. \ref{expansion}(a) by the broad turbulent layer that starts to form in Fig. \ref{DensityEfield}(c) at $r\approx 500 \, \mu m$.

\begin{figure}
\includegraphics[width=1\columnwidth]{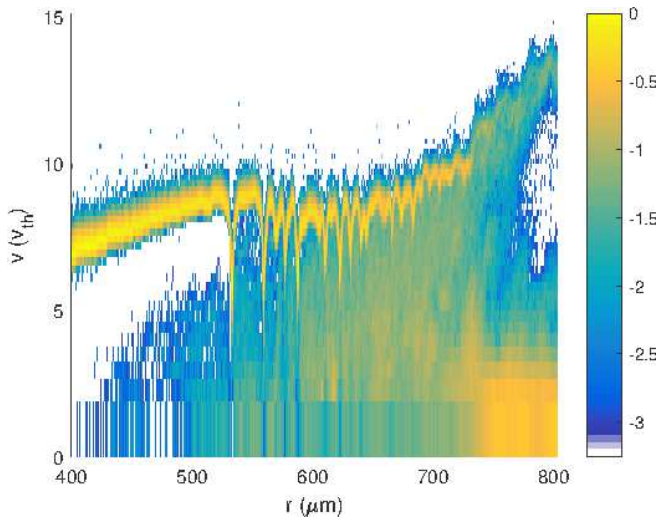}
\caption{Slice of $f_p(r,\alpha=45^\circ,(|v|/v_{th}))$ at 530 ps: the phase space density is normalized to its peak value in the displayed interval and the color scale corresponds to 10-logarithmic phase space density.}
\label{Solitons}
\end{figure}

Figure \ref{Solitons} shows $f_p(r,\alpha,(|v|/v_{th}))$ for $\alpha = 45^\circ$ and $t=530$ ps. 
The blast shell protons enter with the speed $8 \times 10^5$ m/s at $r=450 \, \mu m$ and are slowed down by the ISW at $r \approx 530\, \mu m$, which is the density band to the left in Fig. \ref{DensityEfield}(f). The amplitude of its velocity modulation is $4v_{th}$, which is close to $c_s$, and its width is about 5 electron Debye lengths $\lambda_D \equiv {(\epsilon_0 k_B T_0/n_0e^2)}^{1/2}=1.35 \mu m$ of the ambient plasma. The large amplitude of the ISW is close to the limit, at which it changes into a shock \cite{Malkov15}. 

The ISW has moved from $r\approx 410\mu m$ in Fig. \ref{DensityEfield}(d) to $\approx 530\mu m$ in Fig. \ref{DensityEfield}(f) and its speed $3.6\times 10^5$ m/s is approximately constant (See Fig. \ref{expansion}). The ISW propagates at the speed $\approx c_s$ towards lower values of $r$ in the rest frame of the proton beam that moves with the speed $\sim 8 v_{th}$ in Fig. \ref{Solitons}. The local ion acoustic speed exceeds $c_s$ because the electron temperature, averaged over the interval $525 \mu m < r < 540 \mu m$ in which the ISW is located, is 30\% larger than $T_0$ (not shown). Even the strongest ISW is thus not an IAS, which would require it to propagate faster than the local $c_s$ \cite{Tran79}. 


The blast shell protons traverse the ISW and encounter a second one at $r \approx 560 \, \mu m$. More ISW's are observed to the right, which form the entangled ISW's in Fig. \ref{DensityEfield}(f). The size of the ISW's and the density of the blast shell protons decreases with each ISW crossing and the hot proton background gets denser. The hot low-energetic protons have a density minimum at the location of each ISW; the electric potential of each ISW repels protons. 

In conclusion we have modeled the expansion of a radial blast shell into a uniform plasma. A shock formed, which moved at more than twice the ion acoustic speed and compressed, heated and accelerated the ambient protons. ISW's formed in the post-shock plasma, which consisted of a dense beam of blast shell protons and the shock-heated ambient protons. The ISW's grew in a turbulent plasma and, hence, they were non-planar to start with. An instability amplified their initial oscillations. The electric field distributions of these entangled ISW's and their interaction with the shocked ambient protons slowed down and compressed the blast shell protons and helped confining the shock-heated ambient protons. 
 
M. E. D. acknowledges financial support by a visiting fellowship of CRAL. M. B. and G. S. acknowledge financial support by the EPSRC grants: EP/P010059/1 and EP/K022415/1. 
The simulations were performed on resources provided by the Grand Equipement National de Calcul Intensif (GENCI) through grant x2016046960.

\end{document}